\documentclass[aps,prl,twocolumn, amsmath, amssymb]{revtex4-1}
\usepackage{color}
\usepackage{graphicx}
\usepackage{siunitx}
\usepackage{todo}
\usepackage{cleveref}

	\definecolor{mgray}{gray}{0.6}
\begin{document}
%\title{Initiation of Crystal Growth Rims}
\title{Cavity formation in confined growing crystals}
\author{Felix Kohler$^1$}
\email[Felix Kohler: ]{felixkohler@gmail.com} % 
%\affiliation{Physics of Geological Processes, University of Oslo, Oslo, Norway}
%\email[Felix Kohler: ]{felix.kohler@fys.uio.no} %This address will probably not exist for very long
% * <olivier_pierrelouis@yahoo.fr> 2017-11-07T07:42:17.568Z:
%
% ^.

\author{Luca Gagliardi$^2$}
\email[Luca Gagliardi: ]{luca.gagliardi@univ-lyon1.fr}
%\affiliation{Institut Lumi\`ere Mati\`ere, UMR5306 Universit\'e Lyon 1-CNRS, Universit\'e de Lyon 69622 Villeurbanne, France}

\author{Olivier Pierre-Louis$^2$}
%\email[Olivier Pierre-louis: ]{olivier.pierre-louis@univ-lyon1.fr}

\author{Dag Kristian Dysthe$^1$}
%\email[Dag Kristian Dysthe: ]{d.k.dysthe@fys.uio.no}
%\affiliation{Physics of Geological Processes, University of Oslo, Oslo, Norway}

\affiliation{$^1$ Physics of Geological Processes, University of Oslo, Oslo, Norway}
\affiliation{$^2$ Institut Lumi\`ere Mati\`ere, UMR5306 Universit\'e Lyon 1-CNRS, Universit\'e de Lyon 69622 Villeurbanne, France}

\date{\today}
\pacs{} % insert suggested PACS numbers
%\keywords{} % insert suggested keywords - APS authors don't need to do this 

% \textcolor[rgb]{1,0,0}{[Abstract: no more than 600 characters, including spaces]}
%----------ABSTRACT-------------------
\begin{abstract} 
Growing crystals form a cavity when placed against a wall.
The birth of the cavity is observed both by
optical microscopy of sodium chlorate crystals (NaClO$_3$) growing in the vicinity of a glass surface,
and in simulations with a thin film model.
The cavity appears when  growth cannot be 
maintained in the center of the contact region due to an insufficient supply 
of growth units through the liquid film between the crystal and the wall.
We obtain a non-equilibrium morphology diagram characterizing the conditions
under which a cavity appears. 
Cavity formation is a generic phenomenon at the origin of the formation
of growth rims observed in many experiments, and is a source
of complexity for the morphology of growing crystals in natural environments.
Our results also provide restrictions for the conditions under which
compact crystals can grow in confinement.

\end{abstract}
\maketitle

In natural environments, confinement commonly constrains the growth of crystals~\cite{Alba-Simionesco2006c}. 
Constrained growth may cause large forces such as in salt weathering~\cite{Goudie1997,Rijniers2005,Flatt2014},
in the opening of veins in the Earth's crust~\cite{Wiltschko2001,Gratier2012a},
or in frost heave~\cite{Wilen1995,Rempel2001}.
In biomineralization 
--the process by which living organisms grow minerals,
confinement also plays a key role 
to control the shape and phase of nano-crystals~\cite{Cantaert2013,Wang2014},
and combines with the chemical environment~\cite{Tanaka2011}
to govern microstructure formation in, e.g., bones or dentine.
Beyond its relevance for natural environments,
motion produced by confined growth can be used in technological
applications such as nanomotors~\cite{Regan2005}.
However, while the morphology of freely growing crystals has been investigated
for decades~\cite{Saito1996,Pimpinelli1998}, much less is known about crystal morphological evolution in confinement.
Here, we show that the simplest confinement, i.e., the vicinity of a flat
impermeable substrate, leads to the formation of a cavity in the growing crystal.
The cavity forms  due to insufficient material supply 
in the center of the contact.
After their formation, cavities can expand up to the edge of the
contact, leading to growth rims that have been observed
in force of crystallization experiments since the beginning of the 20th century~\cite{Becker1905, Becker1916, Weyl1959,Royne2012}.

Cavity formation is observed both using 
optical microscopy of sodium chlorate crystals (NaClO$_3$) growing in the vicinity of a glass surface,
and in simulations based on a thin film model.
The birth of the cavity is characterized by a non-equilibrium morphology diagram
describing the balance between growth rate and mass supply.
This diagram found to be robust with respect to
variations in the properties of the growth mechanism
such as anisotropy or kinetics, and therefore provides generic 
conditions for growing compact crystals without cavities in
micro and nano confinement conditions such as those encountered in
the Earth's crust, in biomineralization, or in technological applications.

\paragraph{\bf Experimental methods and observation of the cavity.}

\begin{figure}
	\centering
        		\includegraphics[width=1.00\linewidth]{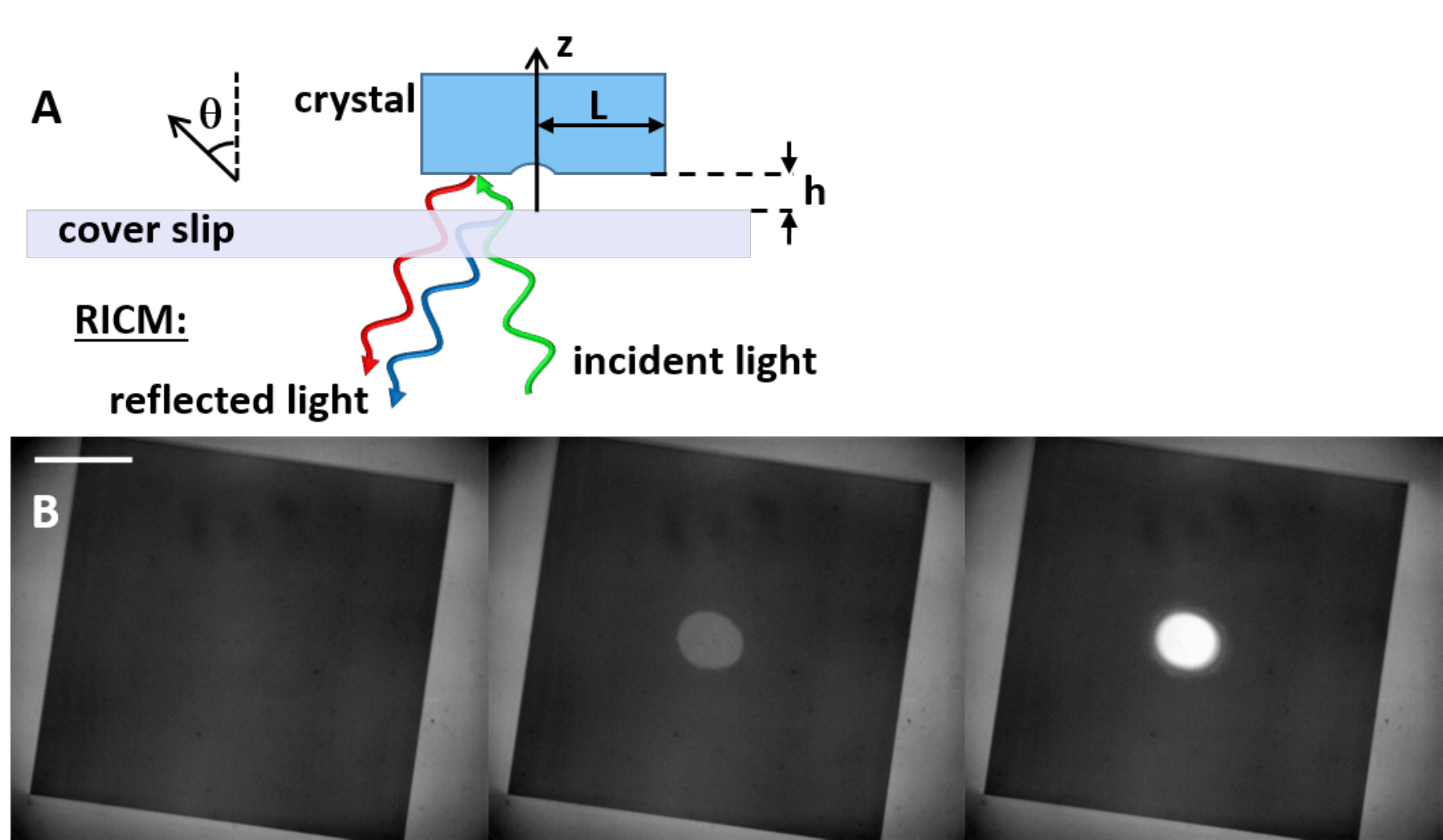}

	\caption{\textbf{A:} Experimental setup and observations.
    A growing crystal is placed against a glass substrate.
    The crystal surface profile is determined with $nm$ accuracy
    by RICM using the interference between the light reflected by the crystal 
    interface (shown in red)  with the light reflected by the glass-solution interface (shown in blue). 
    \textbf{B:} RICM images showing the formation of a cavity as growth proceeds.
    Snapshots just before the start of cavity formation, 15min later, and 35min later.
    Crystal size: 188$\mu m$ x 192$\mu m$, supersaturation: $\sigma_b=0.093$, distance to the glass substrate: $h=51nm$.
    }
	\label{fig:setup}
\end{figure}
In our experiments, we control the solution supersaturation while measuring the confined crystal topography. 
A  NaClO$_3$ seed crystal with a volume of $\sim 1$~mm$^3$ is placed in a 60 $\mu$l chamber filled with a saturated NaClO$_3$ solution. 
The solubility $c_0(T)$ of NaClO$_3$ is strongly temperature dependent~\cite{seidell1940, Crump1995, Kerr1985}.  
The temperature of the sample chamber and oil immersion objective is controlled with a long term precision of 1~mK. By adjusting the temperature $T$ below or above the equilibrium temperature $T_{eq}$,
to obtain growth or dissolution, the relative saturation $\sigma_b =\big(c_b-c_0(T)\big)/c_0(T)$ can be controlled with an accuracy of 0.1\%.
The equilibrium point, $c_b=c_0(T_{eq})$, is identified when the crystal exhibits roundish edges and neither grows nor dissolves. 
The high nucleation barrier of NaClO$_3$ prevents the appearance of other seed crystals in the chamber that could affect the concentration of the bulk solution~\cite{Qian1998}.

The confined crystal interface is observed from below using reflection interference contrast microscopy (RICM), 
which is based on the interference between reflections from the
glass interface and the confined crystal interface (see fig.\ref{fig:setup}). 
Using a specialized objective, a high power LED light source and a 16bit camera this method
allows us to determine the distance $\zeta(\bf{r})$ between the crystal and the glass with nm precision~\cite{Limozin2009a}. 
Due to the presence of dust grains on the substrate, 
the distance $\zeta(\mathbf r)$ cannot be decreased below a minimum value, which ranges from 10nm to 80nm.
In order 
to gain more control on the gap between the substrate and the crystal we 
have also performed experiments with  glass beads  deposited 
on the substrate prior to the seed crystal, which act as calibrated spacers (see SM~\cite{SUP} for details).
In all measurements, the lateral extent $2L$ of the crystal facet facing the substrate is determined by tracking the edges with a precision of 15nm.

Our main observation is that during growth when the size $2L$
exceeds a critical value, which depends on 
the average film width $h$  and  on the supersaturation $\sigma_b$, 
a cavity forms within the contact region. 
Snapshots of the temporal evolution 
slightly above the threshold are shown in \cref{fig:setup}.
The corresponding surface plots
are shown in \cref{fig:cav}A.
See also the corresponding movie in Supplemental Material (SM)~\cite{SUP_expVideo}. 
The appearance of the cavity can be interpreted 
as a consequence of a lower growth rate in the central part of the facet as compared to
the parts closer to the facet edges. Intuitively, this lower growth rate is due to confinement limiting the diffusive mass supply from the bulk liquid. 
In order to assess 
the influence of material properties and physical conditions
on cavity formation,  we have performed numerical simulations of a thin film model
describing the dynamics within the contact region.
\begin{figure*}
    \includegraphics[width=1.0\textwidth]{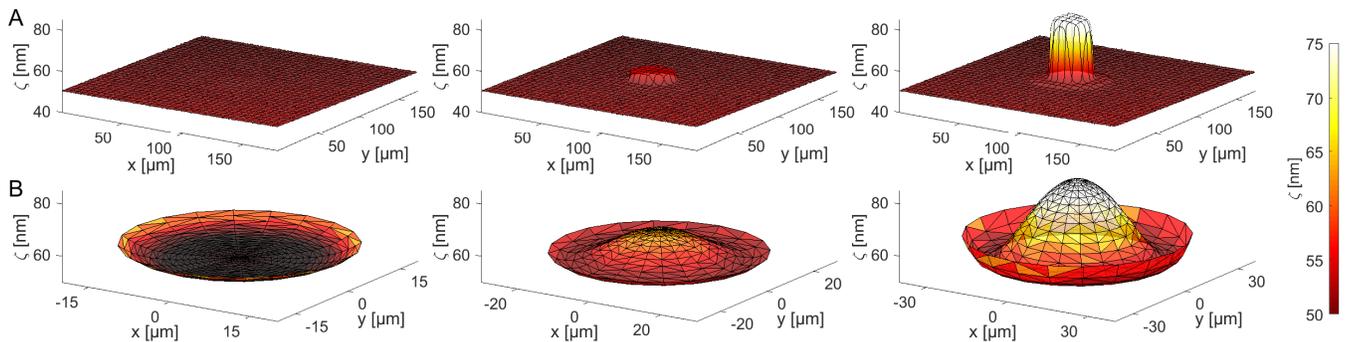}
\caption{3D view of cavity formation at the confined crystal interface.
\textbf{A}: %Corresponding 
Surface plot of the distance $\zeta(\bf{r})$ between crystal interface and glass substrate from the RICM images
reported in \cref{fig:setup}B.
\textbf{B}: Simulation result showing axisymmetric steady states 
with film width $h = 50 \mathrm{nm}$,
supersaturation $\sigma_{BC}\approx 0.004$ and thickness $\zeta_{BC}=1040$nm at the edge of the simulation box of radius $R$. 
The surface plots represent only the contact region of radius $L<R$ with supersaturation $\sigma_b=\sigma(L)<\sigma_{BC}$ at their edge. 
From left to right :  
$R=60 \mathrm{\mu m}$, $L\approx 20 \mathrm{\mu m}$, and $\sigma_b \approx 0.0011$; 
$R=65 \mathrm{\mu m}$, $L\approx 28 \mathrm{\mu m}$, and $\sigma_b \approx 0.0014$; 
$R=70 \mathrm{\mu m}$, $L\approx 37 \mathrm{\mu m}$, and $\sigma_b \approx 0.0019$.
}
\label{fig:cav}
\end{figure*}

\paragraph{\bf Simulation model and observation of the cavity.}
The model, based on that of Ref.\cite{Gagliardi2017}, accounts for the dynamics
of growth and dissolution of a crystal, considered as a rigid body without elastic deformation,
coupled to diffusion and hydrodynamics in the liquid film. 
We used some additional simplifying assumptions.
First,  attachment-detachment kinetics of ions are fast at the surface
of salts such as NaClO$_3$. Since in addition, diffusion limited mass transport along the thin liquid film is
decreased by confinement, we can safely assume that kinetics are limited by diffusion in the liquid.
Moreover, whereas the bulk solution surrounding the crystal is
influenced by solutal buoyancy convection, which originates from  temperature and concentration gradients, 
such effects can be excluded in the confined solution below the crystal.
Furthermore, we neglect the hydrodynamic flow induced 
by the density difference between the crystal and the solution during growth~\cite{Wilcox1993}, 
though, we keep the density difference as the origin of the gravitational force
$F_z$ maintaining the crystal on the substrate.  The model is axisymmetric about the $\mathbf{z}$ axis
defined in \cref{fig:setup}.

Using the small slope limit and the dilute limit~\cite{Gagliardi2017}, we then obtain two equations
accounting for the evolution of the local thickness $\zeta(r,t)$ of the liquid film
where $r$ is the radial coordinate, 
and for 
the 
%translational rigid body velocity 
growth rate $u_z(t)$, which is the velocity of the crystal along  $\mathbf{z}$.
The first equation accounts for local mass balance, the second for global force balance
\begin{subequations}
\label{eq:evolution}
\begin{align}
&\partial_t \zeta = -B\frac{1}{r}\partial_r\Bigl[
r\zeta \partial_r(\tilde{\gamma} \partial_{rr} \zeta +\frac{\tilde{\gamma}}{r}\partial_r \zeta  - U'(\zeta)) 
\Bigr] - u_{z}\, ,
\label{eq:h}\\
&u_{z} \, 2\pi\int_0^R \!\!\!\!\mathrm{d}r\, r\int_r^R \!\!\!\!\mathrm{d}r'\, \frac{6\eta r'}{\zeta(r')^3} 
= F_{z} + 2\pi\int_0^R \!\!\!\! \mathrm{d}r \, r\, U'(\zeta)\, .
\label{eq:u}
\end{align}
\end{subequations}
Here, 
 $U(\zeta)$ is the interaction potential between the substrate and the crystal,
$\eta$ is the liquid viscosity, 
and $B=\Omega^2 D c_0/(k_BT)$ is an effective mobility, which combines the diffusion constant  $D$, 
 molecular volume $\Omega$,  numerical solubility $c_0$,  Boltzmann constant $k_B$, and  temperature $T$.
 In addition, we have defined the surface stiffness $\tilde\gamma=\gamma(0)+\gamma''(0)$,
 where $\gamma(\theta)$ is the surface free energy and the
 angle $\theta$ is defined in \cref{fig:setup}.
In order to mimic the experimental conditions where thicknesses smaller than $h$
are forbidden by dust grains, we consider the repulsive potential
\begin{equation}
\label{eq:potential}
U(\zeta) =  {\cal A}\; { f}\Bigl (\frac{\zeta-h}{\bar{\lambda} h}\Bigr )\, ,
% U(\zeta) =  \bar{a}A\frac{e^{\frac{-(\zeta-h)}{h\bar{\lambda}}}}{\zeta-h}\, ,
\end{equation}
where ${\cal A}$ and $\bar{\lambda}$ are constants, and ${f}(x)={\rm e}^{-x}/x$ 
is a Yukawa-like term.

In experiments, the crystal surface facing the substrate is a facet,
and surface stiffness is expected to diverge for faceted orientations~\cite{Saito1996}, 
leading to a singular crystal shape.
Such singularities cannot be handled by our continuum model where the 
crystal shape always exhibits a smooth profile. However, we approach 
the facet behavior by artificially increasing the stiffness to $\tilde\gamma = 10^2 \mathrm{J/m^2}$,
i.e., roughly $10^3$ times larger than the expected surface tension $\gamma(0)\sim 0.1\mathrm{J/m^2}$.
Choosing the other model parameters in a way which is consistent
with the literature and with experiments (see SM~\cite{SUP}), this \textit{ad hoc} assumption on the stiffness allows one
to obtain growth rates and supersaturations comparable to those observed in experiments.

We numerically solved \cref{eq:evolution}(a,b) in a circular simulation box of fixed radius $R$, 
with fixed film width ${\zeta}(R)={\zeta}_{BC}$ and supersaturation $\sigma(R)=\sigma_{BC}$
at the boundary of the integration domain. All simulations were started with a flat contact region.

Steady-state profiles are reached at long simulation times. 
They are reported in \cref{fig:cav}B
for increasing sizes $R$ of the simulation box. 
A movie of the related time evolution is reported in Supplemental Material (SM)~\cite{SUP_simVideo}. 
As in the experiments, we find that a cavity forms
when the size of the crystal exceeds a critical value.
As shown in \cref{fig:cav}B, the effective radius $L$ of the contact is smaller than
the total radius $R$ of the simulation box.
Despite the absence of growth-induced expansion of the contact size $L$ in simulations,
good qualitative agreement is obtained with the experiments. This agreement suggests 
a quasistatic behavior, where the evolution of the lateral crystal size is slow enough
to have a negligible influence on the diffusion field in the contact region.

\paragraph{\bf Criterion for cavity formation.}
Based on this hypothesis of quasistatic dynamics,
the threshold for cavity formation can be deduced from global mass conservation.
Within the thin film approximation, 
the concentration does not depend on the $z$ coordinate, 
and mass balance for a disc of radius $r$  and constant thickness $h$ of liquid film centered
in the contact region reads
\begin{equation}
\pi r^2 J_k = - 2 \pi r h J_d(r), 
\label{eq:flux}
\end{equation}
where $J_k=u_z/\Omega$ is the mass flux entering the crystal per unit facet area,
and $J_d(r)=-D(dc/dr)$ denotes the diffusion flux entering into the liquid volume. 
The concentration is integrated as
\begin{equation}
c(r)= c_b- \frac{J_k}{4 h D}\left(L^2-r^2\right) ,
\label{eq:cx}
\end{equation}
where $c_b$ is the concentration at the edge of the contact region.
The local supersaturation
$\sigma(r)=c(r)/c_0-1$ decreases toward the center of the facet.
We expect that growth can be maintained in the central region only if the supersaturation
is positive at $r=0$. 
This is confirmed by the numerical solution of \cref{eq:evolution} 
showing that a cavity starts forming approximately when the 
supersaturation vanishes in the center of the contact.
We therefore obtain a condition for cavity formation from the condition $\sigma(0)\leq 0$, which can be rewritten as
\begin{equation}
u_z \geq 4 \Omega c_0 \sigma_b D  \frac{h}{L^2}= \beta D,
\label{eq:ccrit}
\end{equation}
where $\beta= 4 \Omega c_0 \sigma_b {h}/{L^2}$, and $\sigma_b=\sigma(L)=c_b/c_0-1$.

\begin{figure}
\includegraphics[width=\columnwidth]{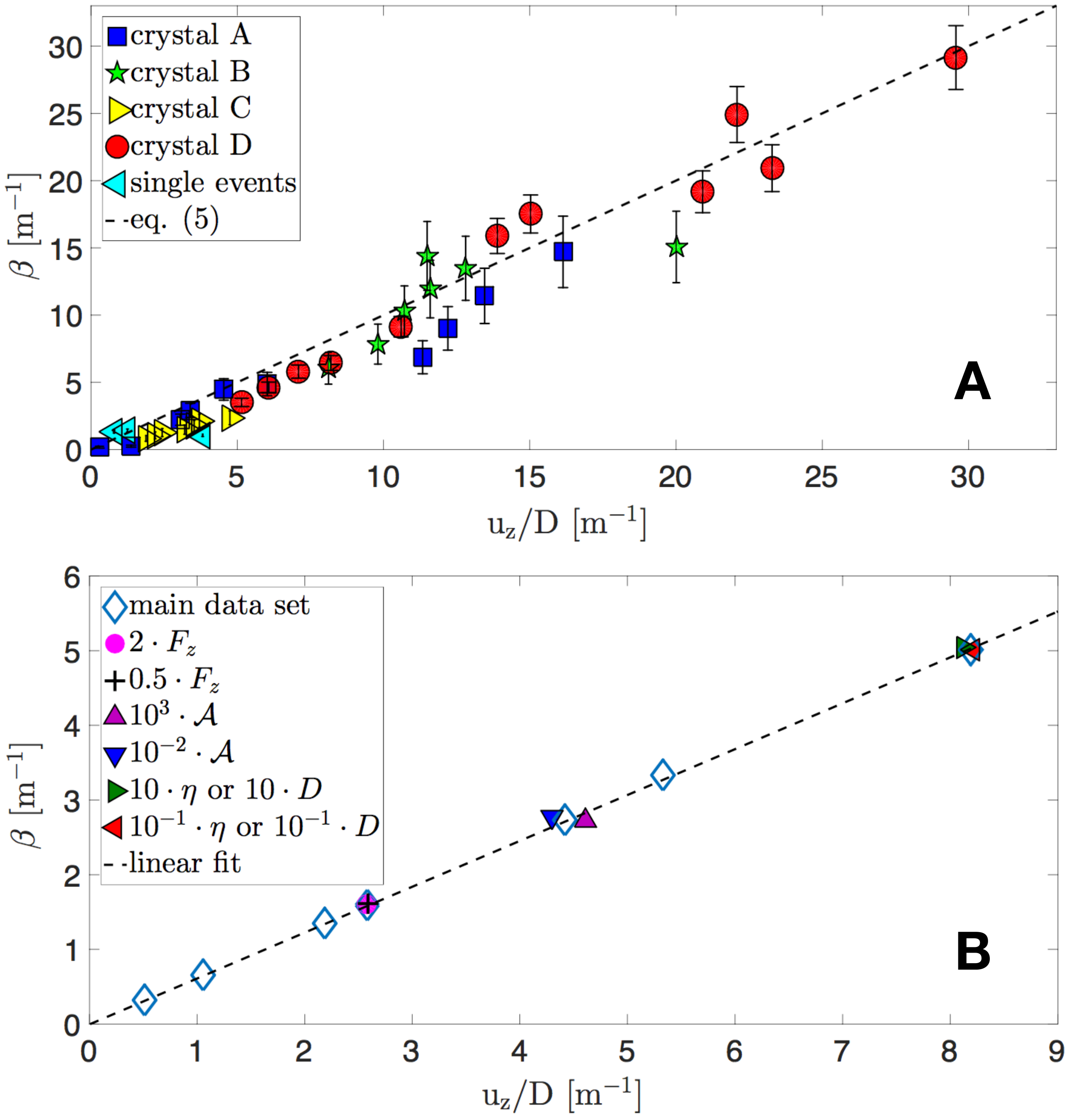}
\caption{ Non-equilibrium  morphology diagram for cavity formation.
\textbf{A}  Transition line in experiments for different crystals. 
Results plotted assuming $D = 0.0935\times 10^{-9}$\si{m^2s^{-1}} which allows for
a perfect correspondence with \cref{eq:ccrit}.
\textbf{B} Transition line obtained from simulations. 
Colored filled dots were obtained using different values of the repulsion strength $\bar{a}$, 
viscosity $\eta$ or diffusion $D$ and external force $F_z$, with respect to the main set of simulations.
}
\label{fig:phase}
\end{figure}

\paragraph{\bf Simulation morphology diagram.}
Simulation results reported in the $(u_z/D,\beta)$ plane in \cref{fig:phase}B, indeed reveal
a linear behavior of the transition line between
the flat regime and the cavity regime,
as predicted by \cref{eq:ccrit}.
However, the slope $D\beta/u_z\approx 0.61$ is slightly lower than the expected value $D\beta/u_z=1$.
Details on the methods to determine the transition point,
the contact size $L$, and supersaturation $\sigma_b=\sigma(L)$ 
at the edge of the contact region are reported in SM~\cite{SUP}.
We have checked that these conclusions are not affected by the boundary
conditions imposed for numerical integration. 
Since the quantity $\beta$ depends on $h$, the dimensionless range $\bar \lambda$ of the repulsion potential
is kept very small $\sim 10^{-2}$ so that the liquid film thickness in the stable regime is 
approximately equal to $h$ in all simulations.

One striking property of the transition line is its robustness 
with respect to the variation of the physical parameters
that do not enter into \cref{eq:ccrit}.
Indeed, as shown in \cref{fig:phase}B, 
large variations in gravitational force $F_z$,
and normalized interaction amplitude ${\cal A}$, lead to negligible 
changes in the transition line position.
Furthermore, increasing or decreasing one of the kinetic constants
$D$ or $\eta$ by a factor of 10 also does not affect 
the transition line. 

\paragraph{\bf Experimental morphology diagram.}
In order to explore the transition in experiments,
we have performed growth cycles. 
This procedure allowed us to explore a range of supersaturations with a single sample.
For each cycle, we monitored the surface profile $\zeta(\bf{r})$ 
during growth at fixed supersaturation, and recorded the
critical size at which the cavity forms.
As soon as the depth of the cavity exceeded $15$nm, 
the temperature was increased to attain a saturation value at which the cavity closes again.
Once we obtained a flat interface, the entire procedure 
was automatically repeated with a different growth supersaturation.
The vertical growth rate $u_z$ can be obtained
from the increase of the depth of the cavity just after its formation.
This method assumes that the growth rate at the bottom of the cavity is negligible
leading to a deepening which is only due to the growth rate $u_z$ of the 
contact region outside the cavity. 
% {\color{red} [[Felix: I do not see the issue with the $u_z$ measurement. For sure, it is not the most precise. However, it is a quite direct measurement, worked better than expected and was performed over large range of growth rates. I would not mention the layer counting experiment. We would open a completely new chapter - including layer growth and visualization.]]}
Moreover, since the lateral growth rate $u_x$ is easier to determine
than $u_z$ from the growth of the cavity, we measure $u_x$, 
and determine $u_z$ from a linear interpolation
of the relation between the two velocities based on a large number of measurements. The ratio $u_x/u_z$ is roughly independent of $L$, $h$, and $\sigma_b$, as shown in SM~\cite{SUP}.
In addition, geometrical corrections described in SM~\cite{SUP}
are used to evaluate $\beta$ for elongated and inclined crystals.

Our measurements reported in \cref{fig:phase}A agree with
a linear behavior of the transition line in the $(u_z/D,\beta)$ plane.
Diffusion constants $D=0.093 \cdot 10^{-9} $m$^2$s$^{-1}$ or $D=0.057 \cdot 10^{-9} $m$^2$s$^{-1}$
respectively provide quantitative agreement with the slopes predicted 
by \cref{eq:ccrit} or by simulations in \cref{fig:phase}B.
These constants are consistent with values reported in the literature~\cite{Campbell1969a}.

\paragraph{\bf Discussion.}
Interestingly, the dependence of $u_z$ on physical parameters
is different in simulations and experiments. For example, while
$u_z$ is roughly independent of $h$ in experiments, 
 we observe that $u_z$ is proportional to $h$ is simulations.
Despite this difference, both simulations and experiments indicate that the 
transition line is linear in the $(u_z/D,\beta)$ plane.

Additional differences between experiments and simulations
have been observed. First, the shape of the cavity is less rounded
in experiments, and emerges from a flat surrounding facet,
as seen in Fig.\ref{fig:cav}A.
This could be related to the limited description of anisotropy of the model within the 
small slope approximation. Second, close to the threshold,  
random opening and closure of the cavity are observed in experiments. 
The results reported above correspond to the lower boundary of the 
stochastic transition regime.
Such fluctuations could be attributed to a nucleation-like process
associated to the competition between thermal fluctuations, and 
surface tension driven decay of the cavity.
Again, despite these differences, both experiments
and simulations collapse on a 
linear transition line in the  $(u_z/D,\beta)$ plane.
The robustness of this linearity can be traced back to the fact that it depends only 
on  two ingredients: mass conservation, and diffusion-limited mass transport,
as discussed  above in the derivation of \cref{eq:ccrit}.

In conclusion, we have shown that when a growing crystal
is placed in the vicinity of a flat wall, a cavity forms in the 
surface of the crystal facing the wall.
The presence of a cavity can be predicted from the crossing of a linear
transition line in the $(u_z/D,\beta)$ plane. 
Cavity formation in confinement appears as an alternative path toward
the formation of concave crystals, beyond well known 
free growth instabilities
leading, e.g., to dendrites~\cite{Langer1980} or hopper crystals\cite{Sunagawa1990}.

In later stages of growth the cavity can expand up to the edges of the contact area,
ultimately leading to a growth rim, as observed in force of crystallization 
experiments~\cite{Becker1905,Becker1916}. 
Since the birth of the cavity affects the shape and area of the contact,
cavity formation should also influence the force and interactions between
the crystal and its environment.

\begin{acknowledgments}
The authors wish to acknowledge funding from the European Union's Horizon 2020 research and innovation program under grant agreement No 642976.
\end{acknowledgments}
\bibliography{Mendeley_Cavity}
\end{document}